\documentclass[useAMS,usenatbib,psfig]{mn2e}
\usepackage{psfig}
\usepackage{times}

\title[Anisotropic radiation from accretion disc-coronae] {Anisotropic radiation from accretion disc-coronae in active galactic nuclei}
\author[Y.-D. Xu]
{Ya-Di Xu\\Department of Physics and Astronomy, Shanghai Jiao Tong
University, 800 Dongchuan Road, Shanghai 200240, China;
ydxu@sjtu.edu.cn}

\date{submitted to MNRAS}

\pagerange{\pageref{firstpage}--\pageref{lastpage}} \pubyear{}

\begin{document}

\maketitle \label{firstpage}

\begin{abstract}
In the unification scheme of active galactic nuclei (AGN), Seyfert
1s and Seyfert 2s are intrinsically same, but they are viewed at
different angles. However, the Fe K$\alpha$ emission line luminosity
of Seyfert 1s was found in average to be about twice of that of
Seyfert 2s at given X-ray continuum luminosity in the previous work
\citep*[][]{2014MNRAS.441.3622R}. We construct an accretion
disc-corona model, in which a fraction of energy dissipated in the
disc is extracted to heat the corona above the disc. The radiation
transfer equation containing Compton scattering processes is an
integro-differential equation, which is solved numerically for the
corona with a parallel plane geometry. We find that the specific
intensity of X-ray radiation from the corona changes little with the
viewing angle $\theta$ when $\theta$ is small (nearly face-on), and
it is sensitive to $\theta$ if the viewing angle is large
($\theta\ga 40^\circ$). The radiation from the cold disc, mostly in
infrared/optical/UV bands, is almost proportional to $\cos\theta$
when $\theta$ $\la 40^\circ$, while it decreases more rapidly than
$\cos\theta$ when $\theta\ga 40^\circ$ because of strong absorption
in the corona in this case. For seyfert galaxies, the Fe K$\alpha$
line may probably be emitted from the disc irradiated by the X-ray
continuum emission. The observed equivalent width (EW) difference
between Seyfert 1s and Seyfert 2s can be reproduced by our model
calculations, provided Seyfert 1s are observed in nearly face-on
direction and the average inclination angle of Seyfert 2s $\sim
65^\circ$.
\end{abstract}


\section{Introduction}

According to the unification scheme of active galactic nuclei (AGN),
Seyfert 1s (Sy1s) and 2s (Sy2s) are intrinsically same but viewed at
different angles, which leads to different observational features
\citep*[][]{1993ARA&A..31..473A}. \citet{2010ApJ...725.2381L} found
that the Fe K$\alpha$ line luminosities of Compton-thin Seyfert 2
galaxies are in average 2.9 times weaker than their Seyfert 1
counterparts. \citet{2014MNRAS.441.3622R} found that the Fe
K$\alpha$ line luminosities of Sy1s are about twice of those for
Sy2s at a given X-ray continuum luminosity ($10-50$~keV). The reason
is still unclear. One possibility is that such difference is caused
by anisotropic X-ray emission from these sources. Indeed, the
orientation dependence of emission from AGN has been found and
studied for a long time. {\citet{2010MNRAS.408.1598N} explored the
uncertainty of the bolometric corrections of quasars with different
viewing angles based on the accretion disc models of
\citet{2000ApJ...533..710H}, and found that a value of the
bolometric luminosity for a quasar viewing at an angle of $\approx
30^{\circ}$ will result in $\approx 30\%$ systematic error, if the
emission from the quasar is assumed to be isotropic.
\citet{2013MNRAS.435.3251R} analyzed a sample of radio-loud quasars
and found that the quasar luminosity changes with orientation. The
sources viewed in the face-on direction are brighter than the
edge-on sources by a factor of 2-3. \citet{2005ApJ...618L..79Z}
explained the observed anticorrelation between type ${\rm
\uppercase\expandafter{\romannumeral 2}}$ fraction and the X-ray
luminosity in 2$-$10 keV based on the AGN unification model with
only one intrinsic luminosity function for these two types of AGNs.
Recently, \citet{2014ApJ...787...73D} used two observed luminosity
functions to investigate the intrinsic quasar luminosity function
with the correction of a simple projection effect for the
anisotropic emission of accretion disc. They concluded that the
orientation dependence is the most important one among several
potential corrections. They also claimed that more complex model of
anisotropy may strengthen the orientation effect.

The black hole accretion disc-corona model has been widely used to
explain both the thermal optical/UV and the power-law hard X-ray
emission in the spectral energy distributions (SEDs) of active
galactic nuclei (AGNs)
\citep{1979ApJ...229..318G,1991ApJ...380L..51H,1993ApJ...413..507H,{
1993PASJ...45..775N},1994ApJ...436..599S,2002ApJ...572L.173L,2003ApJ...587..571L,2009MNRAS.394..207C,2012ApJ...761..109Y}
{and galactic black hole candidates (GBHC)
\citep{{1998ApJ...505..854E},{1997ApJ...487..747D},{1997ApJ...487..759D},{2001ApJ...560..885N}}}.
Although there are different physical mechanisms proposed in the
previous works for heating the corona and the interactions between
the cold disc and the hot corona, the disc-corona is mostly
structured as a sandwich-like cylindrically symmetric system
\citep{{1990ApJ...358..375B},1991ApJ...376..214B,{1994A&A...288..175M},{1995MNRAS.277...70Z},{1997MNRAS.286..848W},1998MNRAS.299L..15D,1999MNRAS.304..809D,{1999A&A...341..936D},{2000MNRAS.316..473R},
2001MNRAS.328..958M,2002MNRAS.332..165M}. The optically thin and
geometrically thick hot coronae are vertically connected to the both
sides of an optically thick and geometrically thin accretion disc.
It was also suggested that such hot coronae may play an important
role in launching relativistic jets observed in X-ray binaries/AGN
\citep*[e.g.,][]{2002MNRAS.332..165M,2004ApJ...613..716C,2011MNRAS.416.1324Z,2013ApJ...770...31W,2014ApJ...783...51C}.
In the accretion disc-corona system, the observed thermal optical/UV
emission is believed to originate from the blackbody radiation of
the thin disc passing through the hot corona. A small fraction of
soft photons are inverse Compton scattered by the hot electrons in
the corona, which contribute to the observed power-law hard X-ray
emission of the system. Moreover, the temperature of the electron in
the transition layer between the corona and the disc experiences a
rapid decrease from $\sim 10^{8-9}$K (the hot corona) to $\sim
10^{4-5}$K (the cold disc), the thermal X-ray line emission may be
produced in such transition zone \citep{{2013ApJ...763...75X}}. In
most previous works, either the cooling rate in the corona or the
spectra from the accretion flow are calculated assuming the corona
to be one parallel-plane. The inverse Compton scattering is often
computed by the Monte Carlo simulation based on the escape
probability method
\citep{1977SvA....21..708P,{2008PASJ...60..399K},2003ApJ...587..571L,2009MNRAS.394..207C}.

Assuming that a fraction of viscously dissipated energy in the disc
is transported into the corona (probably by magnetic fields), we can
calculate the structure of the disc-corona accretion with a set of
equations of the accretion flow, such as, energy equation, angular
momentum equation, continuity equation, and state equation, etc., if
the model parameters are given. Then, the emitted spectra from the
accretion flow can be calculated. In this work, we explore the
angle-dependent emission of the accretion disc-corona system in more
detail, by solving a set of equations describing disc-corona
structure and the equation of radiation transfer in the corona
simultaneously. The change of the spectra with the viewing angle is
explored in detail. {Our results are compared with the X-ray
observations of Sy1s and Sy2s.} The disc-corona model employed in
this work is briefly described in \S 2. The calculation method is
introduced in \S 3. We show the results and discussion in \S 4 and
5.

\section{The disc-corona accretion model}

The accretion disc-corona model used in this work is described in
the previous works
\citep*[][]{2009MNRAS.394..207C,{2013ApJ...763...75X}}. The detailed
model description and calculating approach can be found in
\citet{2009MNRAS.394..207C}. Here we only briefly summarize the main
features of the model. The energy equation of the cold thin disc is
\begin{equation}\label{eqenedisk}
Q_{\rm dissi}^{+}-Q_{\rm cor}^{+}+\frac{1}{2}(1-a_{\rm r})Q_{\rm
cor}^{+}=\frac{4\sigma T^{4}_{\rm disc}}{3\tau},
\end{equation} where,
\begin{equation}\label{eqqdissi}
Q_{\rm dissi}^{+}=\frac{3}{8\pi}\dot{M}\Omega_{\rm
k}^{2}(R)\left[1-\left(\frac{R_{\rm in}}{R}\right)^{1/2}\right],
\end{equation}
is the gravitational power dissipated in unit surface area of the
accretion disc at radius $R$ [where, $\Omega_{\rm k}(R)$ is the
Keplerian velocity, $R_{\rm in}=3R_{\rm S}$, $R_{\rm S}=2GM_{\rm
bh}/c^{2}$ is the Schwarzschild radius for a black hole of mass
$M_{\rm bh}$, and $\dot{M}$ is the mass accretion rate of the black
hole]. The third term in the left side of equation (1) represents
that about half of the power dissipated in the corona is radiated
back into the disc by Compton scattering, and the reflection albedo
of the disc $a_{\rm r}=0.15$ is adopted in the calculations. The
right side of the equation represents the power radiated from the
cold disc via blackbody radiation, where $T_{\rm disc}$ is the
effective temperature in the mid-plane of the disc, and $\tau$ is
the optical depth in vertical direction of the disc. The continuity
equation of the disc is
\begin{equation}\label{continuity}
-4\pi R H_{\rm d}(R)\rho(R)v_{\rm R}(R)=\dot{M},
\end{equation}
where $H_{\rm d}(R)$ is the half thickness of the disc, $\rho(R)$ is
the mean density of the disc, and $v_{\rm R}(R)$ is the radial
velocity of the accretion flow at radius $R$. The state equation of
the gas in the disc is
\begin{equation}\label{state}
p_{\rm tot}=p_{\rm gas}+p_{\rm rad}=\frac{\rho kT_{\rm disc}}{\mu
m_{\rm p}}+\frac{1}{3}aT^{4}_{\rm disc},
\end{equation}
where $\mu=(1/\mu_{\rm i}+1/\mu_{\rm e})^{-1}$, $\mu_{\rm i}=1.23$
and $\mu_{\rm e}=1.14$ are adopted. The half thickness of the disc
$H_{\rm d}$ is
\begin{equation}\label{hdisk}
H_{\rm d}=c_{\rm s}/\Omega_{\rm k}=\frac{\sqrt{p_{\rm
tot}/\rho}}{\Omega_{\rm k}}.
\end{equation}
The energy equation of the corona is
\begin{equation}\label{eqqpluscor}
Q^{+}_{\rm cor}=Q^{\rm ie}_{\rm cor}+\delta Q^{+}_{\rm
cor}=F^{-}_{\rm cor},
\end{equation}
where $Q^{\rm ie}_{\rm cor}$ is the energy transfer rate from the
ions to the electrons via Coulomb collisions \citep*[see equation 11
in][]{2009MNRAS.394..207C} , $\delta$ is the fraction of the energy
directly heats the electron, $F_{\rm cor}^{-}=F^{-}_{\rm
syn}+F^{-}_{\rm brem}+F^{-}_{\rm Comp}$ is the cooling rate in unit
surface area of the corona via synchrotron, bremsstrahlung, and
Compton emissions. {The value of $\delta$ can be as high as $\sim
0.5$ by magnetic reconnection, if the magnetic field in the plasma
is strong \citep[][]{1997ApJ...486L..43B,2000ApJ...529..978B}.
Almost all the power dissipated in the hot corona is radiated away
locally, which means that the radiated power in the corona is
independent of the value $\delta$, and the temperature and density
of the electrons in the corona are almost insensitive with this
parameter \citep*[see][for the discussion]{2009MNRAS.394..207C}.} In
the accretion disc-corona model, the detailed physical mechanism for
generating the energy source and heating the corona is still
unclear, though there are some assumptions of corona heating
processes, such as, magnetic fields reconnection assumed in the
previous works
\citep*[e.g.][]{{1998MNRAS.299L..15D},1999MNRAS.304..809D,
2001MNRAS.328..958M,2002MNRAS.332..165M,2009MNRAS.394..207C,{2013ApJ...763...75X}}.
To avoid the complexity, we introduce a parameter, $f_{\rm cor}$,
the ratio of the power dissipated in the corona, $Q_{\rm cor}^{+}$,
to the gravitational power dissipated in the disc, $Q_{\rm
dissi}^{+}$,
\begin{equation}\label{eqqcorf}
Q_{\rm cor}^{+}=f_{\rm cor}Q_{\rm dissi}^{+},
\end{equation}
in our model calculations.

\section{Radiative transfer in the corona}
We consider a parallel plane geometry of the corona above/below the
disc. The cooling processes in the corona include Compton,
bremsstrahlung, and synchrotron emissions. The incident photons from
the disc at $z=H_{\rm d}$ are assumed to be blackbody radiation. For
simplicity, the electron density and temperature of the corona are
assumed to be constant in the $z$ direction.

\subsection{Radiative transfer equation}
The radiative transfer equation of the corona is,
\begin{equation}
\frac{dI_{\nu}(z,\mu)}{ds}=-I_{\nu}(z,\mu)(\kappa_{\nu}^{\rm
ff}+\kappa_{\rm T})+j_{\nu}^{\rm ff}+j_{\nu}^{\rm C}, \label{eq6}
\end{equation}
where $ds=dz/\mu$, $\mu=\cos\theta$, $\theta$ is the angle of the
photon with respect to the vertical direction of the disc, and
$I_{\nu}(z,\mu)$ is the specific intensity. For the different
absorption and emission processes included in the corona, we have
the combined absorption and emission coefficients,
$\kappa_{\nu}=\kappa_{\nu}^{\rm ff}+\kappa_{\rm T}$, is the total
absorption coefficient including
free-free(bremsstrahlung+synchrotron) absorption and Thomson
scattering, $j_{\nu}=j_{\nu}^{\rm ff}+j_{\nu}^{\rm C}$, is the
emission coefficient (emissivity) including
free-free(bremsstrahlung+synchrotron) emission and Comptonization
emission. We can calculate $I_{\nu}(z,\mu)$ when the structure of
the corona, such as electron temperature $T_{\rm e,cor}(r)$ and
electron density $n_{\rm e,cor}(r)$, are given.

\subsection{Absorption and emission coefficients}
The absorption coefficient of Thomson scattering is $\kappa_{\rm
T}=n_{\rm e,cor}\sigma_{\rm T}$, where $\sigma_{\rm T}$ is Thomson
cross section. The absorption coefficient including bremsstrahlung
and synchrotron processes can be described as
\begin{equation}
\kappa_{\nu}^{\rm ff}=j_{\nu}^{\rm ff}/B_{\nu},
\end{equation}
where $B_{\nu}$ is the blackbody emissivity,
\begin{equation}
B_{\nu}(T)=\frac{2h\nu^{3}/c^2}{\exp{(h\nu/kT)-1}},
\end{equation}
and $j_{\nu}^{\rm ff}=(\chi_{\nu}^{\rm Brem}+\chi_{\nu}^{\rm
Syn})/4\pi$ is the emission coefficient of bremsstrahlung and
synchrotron processes. The bremsstrahlung emissivity
$\chi_{\nu}^{\rm Brem}$ and synchrotron emissivity $\chi_{\nu}^{\rm
Syn}$ are taken from \citet{1995ApJ...452..710N} and
\citet{2000ApJ...534..734M}.

The bremsstrahlung emissivity is given by
\begin{equation}
\chi_{\nu}^{\rm brem}=q_{\rm brem}^{-}\bar{\rm
G}~{\exp}{\left(\frac{h\nu}{kT_{\rm e}}\right)},
\end{equation}
where $\bar{\rm G}$ is the Gaunt factor as in
\citet{1986rpa..book.....R},
\begin{equation}
\bar{\rm G}=\frac{h}{kT_{\rm e}}\left(\frac{3}{\pi}\frac{kT_{\rm
e}}{h\nu}\right)^{1/2}~~~~~~{\rm for}~~\frac{kT_{\rm e}}{h\nu}<1,
\end{equation}
\begin{equation}
\bar{\rm G}=\frac{h}{kT_{\rm
e}}\frac{\sqrt{3}}{\pi}\ln\left(\frac{4}{\zeta}\frac{kT_{\rm
e}}{h\nu}\right)~~~~~~{\rm for}~~\frac{kT_{\rm e}}{h\nu}>1.
\end{equation}
The bremsstrahlung cooling rate per unit volume $q_{\rm brem}^{-}$
consists of electron-ion and electron-electron rates,
\begin{equation}
q_{\rm brem}^{-}=q_{\rm ei}^{-}+q_{\rm ee}^{-}.
\end{equation}
The electron-ion cooling rate is
\begin{equation}
q_{\rm ei}^{-}=1.25n_{\rm e}^{2}\sigma_{\rm T}c\alpha_{\rm f}m_{\rm
e}c^{2}F_{\rm ei}(\theta_{\rm e}),
\end{equation}
where $\alpha_{\rm f}$ is the fine-structure constant, $\theta_{\rm
e}=kT_{\rm e}/m_{\rm e}c^{2}$ is the dimensionless electron
temperature, and the function $F_{\rm ei}$ has the form
\begin{equation}
F_{\rm ei}(\theta_{\rm e})=4\left(\frac{2\theta_{\rm
e}}{\pi^{3}}\right)^{1/2}(1+1.781\theta_{\rm e}^{1.34})~~~~~~{\rm
for} ~~\theta_{\rm e}<1,
\end{equation}
\begin{equation}
F_{\rm ei}(\theta_{\rm e})=\frac{9\theta_{\rm
e}}{2\pi}[\ln(1.123\theta_{\rm e}+0.48)+1.5]~~~~~~{\rm for}
~~\theta_{\rm e}>1.
\end{equation}
The electron-electron cooling rate is
\begin{displaymath}
q_{\rm ee}^{-}=n_{\rm e}^{2}cr_{\rm e}^{2}\alpha_{\rm
f}c^{2}\frac{20}{9\pi^{1/2}}(44-3\pi^{2})\theta_{\rm e}^{3/2}\\
(1+1.1\theta_{\rm e}+\theta_{\rm e}^{2}-1.25\theta_{\rm e}^{5/2})
\end{displaymath}
\begin{equation}
~~{\rm for}~~\theta_{\rm e}<1,
\end{equation}
\begin{equation}
q_{\rm ee}^{-}=n_{\rm e}^{2}cr_{\rm e}^{2}\alpha_{\rm
f}c^{2}24\theta_{\rm e}(\ln1.1232\theta_{\rm e}+1.28)~~{\rm
for}~~\theta_{\rm e}>1,
\end{equation}
where $r_{\rm e}=e^{2}/m_{\rm e}c^{2}$ is the classical electron
radius.

The synchrotron emissivity is given by
\begin{equation}
\chi_{\nu}^{\rm syn}=4.43\times 10^{-30}\frac{4\pi n_{\rm
e}\nu}{K_{2}(1/\theta_{\rm e})}I^{\prime}(\frac{4\pi m_{\rm
e}c\nu}{3eB\theta_{\rm e}^{2}}),
\end{equation}
and the function $I^{\prime}(x)$ is given by
\begin{equation}
I^{\prime}(x)=\frac{4.0505}{x^{1/6}}(1+\frac{0.4}{x^{1/4}}+\frac{0.5316}{x^{1/2}})\exp(-1.8899x^{1/3}),
\end{equation}
where $K_{2}$ is the second order modified Bessel function, $B$ is
the magnetic field strength. For the equipartition magnetic field
case, we have $B=\sqrt{8\pi p_{\rm gas,cor}}$ ($p_{\rm gas,cor}$ is
the gas pressure in the corona).

The emissivity of the inverse Compton scattering is calculated with
the method proposed in \citet{1990MNRAS.245..453C},
\begin{equation}\label{eqjc}
j_{\nu}^{\rm C}=\frac{f_{\nu}(T_{\rm e},n_{\rm
e},I_{\nu^{\prime}})}{4\pi},
\end{equation}
where
\begin{equation}\label{eqiseed}
I_{\nu^{\prime}}=\int_{-1}^{1}{I_{\nu^{\prime}}(z,\mu)\kappa_{\rm T}
2\pi d\mu},
\end{equation}
is the seed photon intensity of the unit volume including incident
photons from all the directions.

\subsection{Numerical solution to the radiative transfer equation}

The radiative transfer equation (\ref{eq6}) is an
integro-differential equation. For a given disc-corona structure
(including temperatures and densities of the electrons and ions,
half thicknesses of the disc $H_{\rm d}$ and thicknesses of the
corona $H_{\rm cor}$, etc.), one can solve this equation by
iterations \citep*[e.g.,][]{1998A&A...330..464C}. An initial
solution can be obtained by solving the equation neglecting the
Compton scattering term, i.e.,
\begin{equation}
\frac{dI_{\nu}(z,\mu)}{ds}=-I_{\nu}(z,\mu)(\kappa_{\nu}^{\rm
ff}+\kappa_{\rm T})+j_{\nu}^{\rm ff}.
\end{equation}
We solve the above equation numerically for the corona
with the boundary condition at the disc surface,
\begin{equation}
I_{\nu}(z=H_{\rm d},\mu)=B_{\nu}(T_{\rm d}^{\rm s}),
\end{equation}
in the range $0<\mu<1$, where $T_{\rm d}^{\rm s}$ is the temperature
of the disc surface, and the boundary condition at the upper surface
of the corona,
\begin{equation}
I_{\nu}(z=H_{\rm d}+H_{\rm cor},\mu)=0,
\end{equation}
in the range $-1<\mu<0$. With derived initial solution, we calculate
the emissivity of the inverse Compton scattering from equations
(\ref{eqjc}) and (\ref{eqiseed}), and solve equation (\ref{eq6})
numerically. With derived $I_{\nu}(z,\mu)$, the emissivity of the
Compton emission can be re-calculated with equations (\ref{eqjc})
and (\ref{eqiseed}). We find that the final solution can be achieved
after several iterations when the solutions converge.

Integrating the intensities over the different directions
($\mu=\cos\theta$) and frequencies $\nu$, the energy loss via
radiation from the upper and lower surfaces of the corona can be
calculated by
\begin{displaymath}
Q_{\rm cor}^{\rm rad}=\int_{\nu}d\nu\int_{0}^{1}{I_{\nu}(z=H_{\rm
d}+H_{\rm cor},\mu)2\pi \mu
d\mu}
\end{displaymath}
\begin{equation}\label{eqqcorrad}
+\int_{\nu}d\nu\int_{-1}^{0}{I_{\nu}}(z=H_{\rm d},\mu)2\pi \mu d\mu.
\end{equation}
Subtracting the incident blackbody radiation from the thin disc,
\begin{equation}\label{eqqdiskminus}
Q_{\rm disc}^{-}=\int_{\nu}\pi B_{\nu}d\nu,
\end{equation}
we obtain the cooling rate in unit surface area of the corona,
\begin{equation}\label{eqfcorminus}
F_{\rm cor}^{-}=Q_{\rm cor}^{\rm rad}-Q_{\rm disc}^{-}.
\end{equation}

Given the black hole mass $M_{\rm bh}$, the dimensionless mass
accretion rate $\dot{m}$ ($\dot{m}=\dot{M}/\dot{M}_{\rm Edd}$, where
$\dot{M}_{\rm Edd}=L_{\rm Edd}/0.1c^2$), and the fraction of the
energy directly heats the electron $\delta$, we can derive the disc
structure (such as, the effective temperature in the mid-plane of
the disc, the density in the disc, and the half thickness of the
disc, etc.) as a function of radius $R$ from equations
(\ref{eqenedisk})-(\ref{hdisk}), and equation (\ref{eqqcorf}), when
the ratio of the power dissipated in the corona $f_{\rm cor}$ is
specified. The structure of the corona (such as, temperatures and
densities of the electrons and ions, scaleheight of the corona) can
be derived with equations (\ref{eqqpluscor}) and (\ref{eqfcorminus})
under the assumption of equipartition of the magnetic pressure and
the gas pressure in the corona. We assume the temperature of the
ions in the corona $T_{\rm i,cor}=0.9T_{\rm vir}=0.9GMm_{\rm p}/3kR$
in this work as that in \citet{2009MNRAS.394..207C}. Thus, the
accretion disc-corona structure is available by solving the
radiation transfer equation together with disc-corona equations
described in \S 2.

The specific intensity from the corona at a certain radius $R$ is
obtained as a function of direction and frequency of the photons.
Integrating the intensity over the whole surface of the corona, the
specific luminosity emitted from the corona per steradian is
\begin{equation}\label{eq27}
\frac{dL_{\nu}}{2\pi d\mu}=\int \mu I_{\nu}(z=H_{\rm d}+H_{\rm
cor},\mu) 2\pi rdr.
\end{equation}
Thus, we can obtain the direction-dependent spectrum of an accretion
disc-corona system for a set of given disc parameters.

\section{Results}

We adopt the model parameters as follows, the black hole mass
$M_{\rm bh}=10^8 M_{\odot}$, the maximum radius of the disc-corona
flow $R_{\rm cormax}=100 R_{\rm S}$, and the fraction of energy
directly heated the electron $\delta=0.5$, are fixed in all cases.
The mass accretion rate $\dot{m}$ and the ratio of the power
dissipated in the corona $f_{\rm cor}$ are taken as two free
parameters in our model calculations.


In Figure 1, we show the emergent spectra for four disc-corona
accretion models with different values of parameters $\dot{m}$ and
$f_{\rm cor}$. The values of the two parameters in the models are,
model (a): $\dot{m}=0.1$ and $f_{\rm cor}=0.1$, model (b):
$\dot{m}=0.1$ and $f_{\rm cor}=0.3$, model (c): $\dot{m}=0.5$ and
$f_{\rm cor}=0.06$, and model (d): $\dot{m}=0.5$ and $f_{\rm
cor}=0.1$. For each model, we plot the emergent spectra observed in
five different directions with the lines in different types and
colors. The spectra from $\theta=87^{\circ},84^{\circ}, 76^{\circ},
50^{\circ}$, and $2^{\circ}$ are shown with black dashed, cyan
dotted, magenta dash-dotted, green thin solid, and blue thin
dash-dotted lines, respectively. The corresponding value of $\theta$
is denoted near each spectrum. The red thick solid line is the
spectra integrated over the directions from $\mu=0$ to $\mu=1$. The
four vertical dotted lines represent the four typical frequency
points corresponding to $2500 {\rm \AA}$, 0.1 keV, 2 keV, and 10
keV, respectively. We can see from these figures that the spectra
from different emitting directions are different both in the
luminosity and the spectral shape. The luminosity decreases with the
increasing viewing angle between the emitting photons and the axis
of the accretion disc, $\theta$. The spectral shapes are similar in
almost all bands but very different between the optical/UV ($\sim
2500 {\rm \AA}$) and soft X-ray band ($0.1\sim 1$ keV).

The different values of mass accretion rate $\dot{m}=0.1$ and
$\dot{m}=0.5$ are adopted in Figure 1(a) and Figure 1(d),
respectively, while the values of all other parameters are the same
in these two figures. The larger the mass accretion rate adopted,
the stronger and harder the spectra are. The different values of
$f_{\rm cor}=0.1$ and $f_{\rm cor}=0.3$ are adopted in Figure 1(a)
and Figure 1(b), respectively, while the values of all other
parameters are the same in these two figures. We find that the
change of the spectra with $f_{\rm cor}$ is also evident. The larger
the value of $f_{\rm cor}$ adopted, the harder the spectra are.

{The spectral shapes plotted in the four panels of Figure \ref{fig1}
show a certain degree of degeneracy of the two parameters $\dot{m}$
and $f_{\rm cor}$. Different combinations of these two parameters
may give very similar spectral shapes, but these can be
distinguished by the different luminosity, which is directly
controlled by $\dot{m}$.}

In order to study the change of spectra with the viewing direction,
we calculate some typical quantities of the observed spectra,
including the observed bolometric luminosity $L_{\rm bol}$, the
spectral luminosity at optical/UV band ($\lambda$=2500 $\rm \AA$)
$L_{\rm o}$, and the spectral luminosity at X-ray band ($E=2$ keV)
$L_{\rm x}$. The optical/UV to X-ray power index $\alpha_{\rm ox}$
is defined as
\begin{equation}
\alpha_{\rm ox}=-\frac {\log L_{\nu(2500~{\rm \AA})}/L_{\nu(2~{\rm
keV})}}{\log \nu(2500~{\rm \AA})/\nu(2~\rm{keV})},
\end{equation}
and the X-ray spectral index between 2 keV and 10 keV, $\alpha_{\rm
x}$ (defined as $L_{\nu,{\rm x}}^{(2-10~{\rm keV})}\propto
{\nu}^{-\alpha_{\rm x}}$). The changes of these quantities with the
viewing angles are plotted in Figures 2-5.

In Figure 2, we plot the change of observed bolometric luminosity
with viewing angle. The red solid, green dashed, cyan dotted, and
black dash-dotted lines correspond to the four models (a)-(d),
respectively. The blue thin dashed line indicates a simple relation
as $L_{{\rm bol}}\propto\cos\theta$ representing the area-projection
effect.


In Figures 3 and 4, we plot the changes of observed spectral
luminosities at optical/UV band ($\lambda=2500{\rm \AA}$) and X-ray
(2 KeV) band with the viewing angle. The red solid, green dashed,
cyan dotted, and black dash-dotted lines correspond to the four
models (a)-(d), respectively. The blue thin dashed lines also
represent simple relations as $L_{\rm o}\propto\cos\theta$ and
$L_{\rm x}\propto\cos\theta$. We find that the two spectral
luminosities (optical and X-ray bands) show different changes with
the viewing angle $\theta$. The X-ray luminosity $L_{\rm x}$
decreases more slowly than $L_{\rm x}\propto\cos\theta$, while
$L_{\rm o}$ decreases more rapidly than $L_{\rm
o}\propto\cos\theta$. The X-ray luminosity $L_{\rm x}$ is almost
isotropic when the viewing angle is small (nearly face-on), and it
becomes strongly anisotropic if the viewing angle is large ($\ga
30^\circ-40^\circ$). The optical luminosity $L_{\rm o}$  is almost
proportional to $\cos\theta$ when $\theta$ $\la 30^\circ-40^\circ$,
but decreases more rapidly than $\cos\theta$ when it is viewed at
large angles ($\ga 30^\circ-40^\circ$).

The changes of the optical/UV to X-ray spectral index, $\alpha_{\rm
ox}$, and the X-ray spectral index between 2 keV and 10 keV,
$\alpha_{\rm x}$, with the viewing angle are shown in Figure 5. The
red solid, green dashed, cyan dotted, and black dash-dotted lines
correspond to the four models (a)-(d), respectively. In each model,
$\alpha_{\rm ox}$ decreases very slowly with $\theta$ at small
viewing angles and quickly at large angles nearly edge-on. The
situation is different for $\alpha_{\rm x}$, which is almost
unchanged with $\theta$ in each model.

{The hard X-ray continuum emission is from the coronae above the
discs, and the discs are irradiated by the X-ray photons from the
corona. Our model calculations show that the angle-dependent X-ray
continuum spectra are anisotropic. They deviate from
$\cos\theta$-dependence. The Fe K$\alpha$ lines are probably emitted
from the irradiated discs. In this case, the Fe line emission is
anisotropic, and its angle-dependence follows $\sim \cos\theta$.
Thus, we can calculate the observed equivalent widths as functions
of viewing angle with different values of model parameters. The
result of model (a) is shown with red solid line in Figure 6 .}

{It is still unclear whether part of observed Fe line emission is
from the torus, which is emitted nearly isotropically. We estimate
how the results would be affected by the torus contribution to Fe
line emission by assuming that $x$ per cent of the total Fe line
luminosity is from the torus. The ratio of the equivalent widths of
the Fe line emission viewing at an angle $\theta$ is}
\begin{equation}
\frac{EW(\theta)}{EW(0^{\circ})}=\frac{(100-x)\cos\theta+x}{100}\times\frac{L_{\nu(2~\rm{keV})}(0^{\circ})}{L_{\nu(2~\rm{keV})}(\theta)}.
\end{equation}
{We re-calculate the relative equivalent width of the Fe line for
different values of $x$=50, 20, and 10. The results are shown in
Figure 6 with black dashed, green dotted, and blue dash-dotted
lines, respectively.}


\begin{figure}
\centerline{\psfig{figure=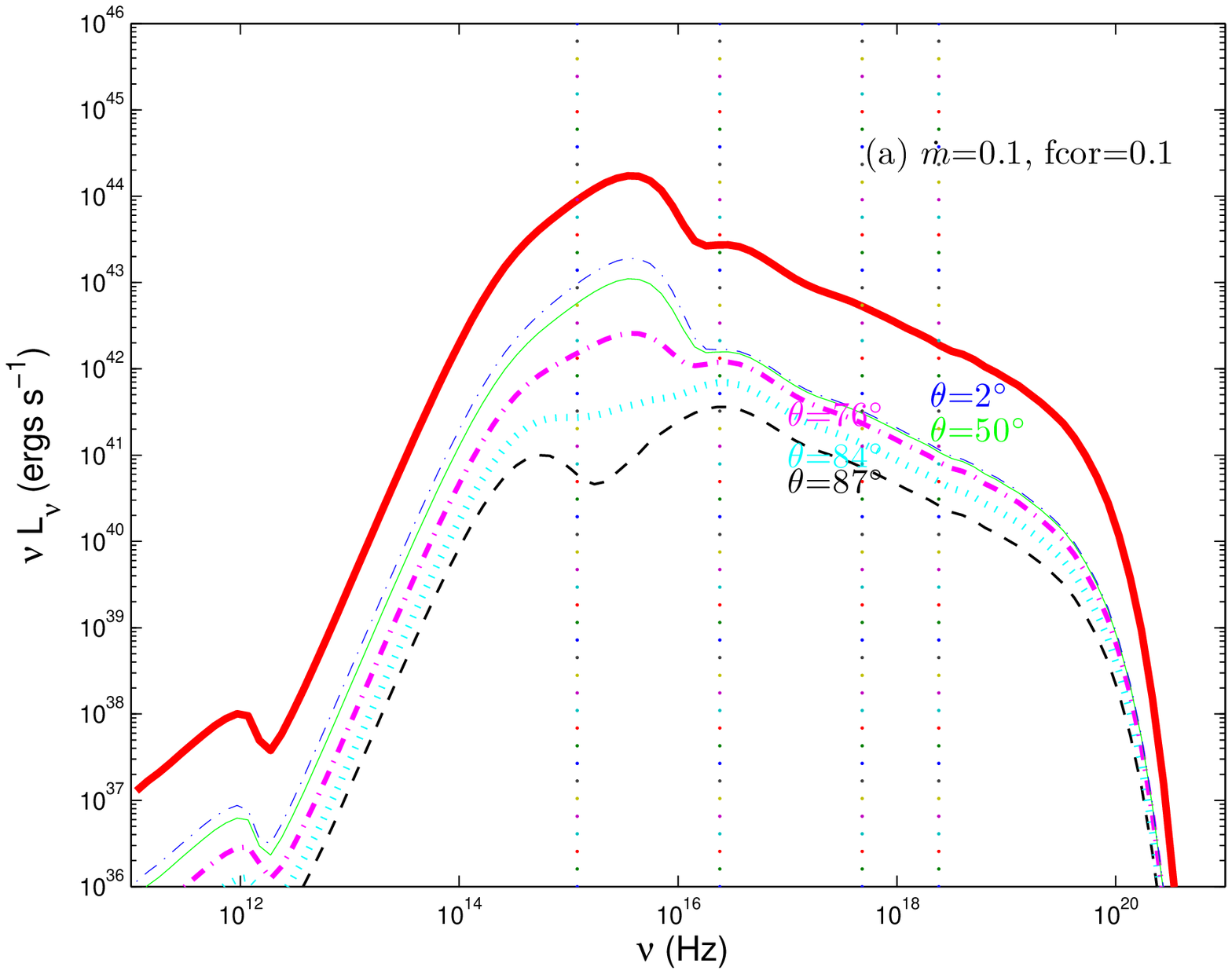,width=7.0cm,height=5.0
cm}}
\centerline{\psfig{figure=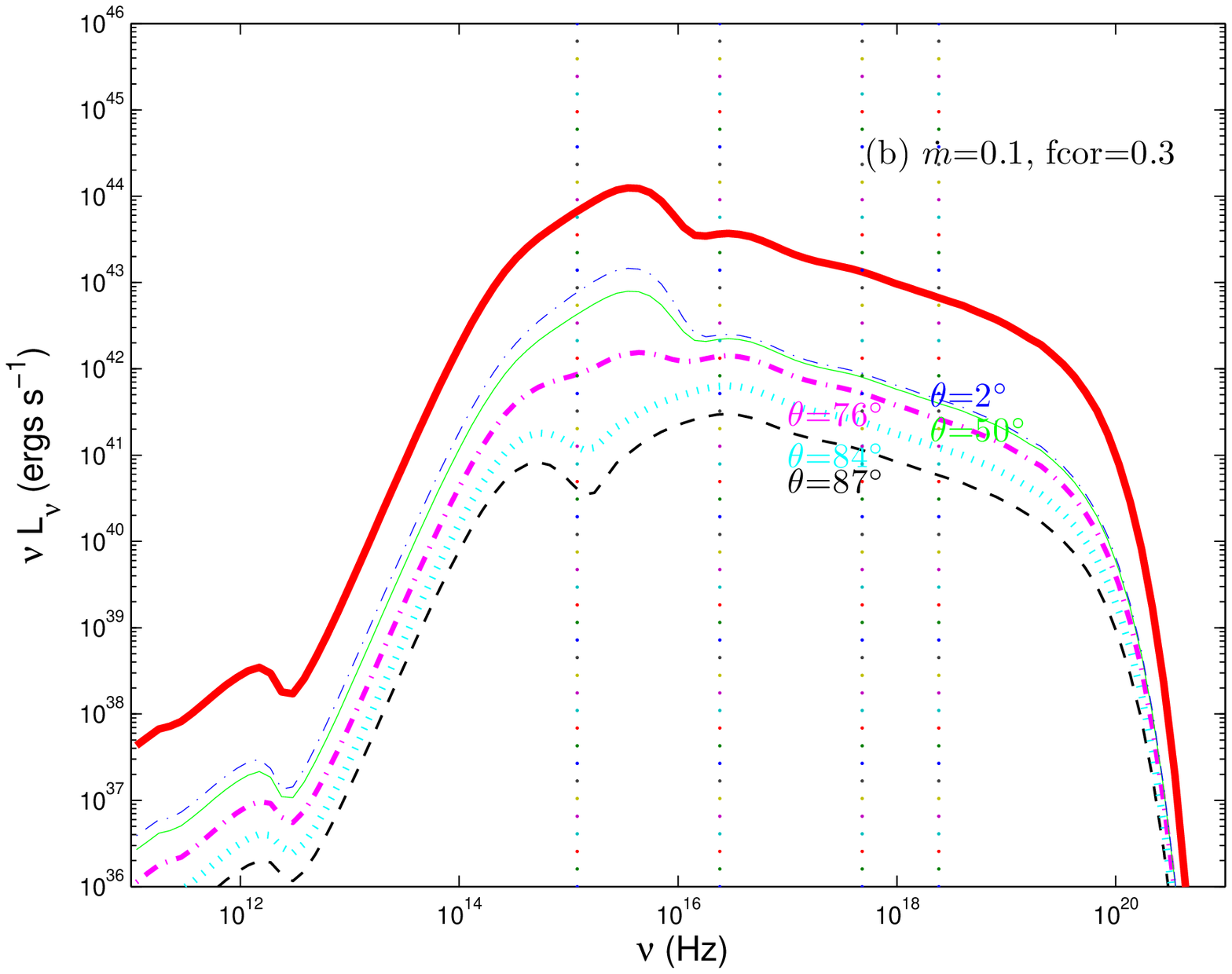,width=7.0cm,height=5.0
cm}}
\centerline{\psfig{figure=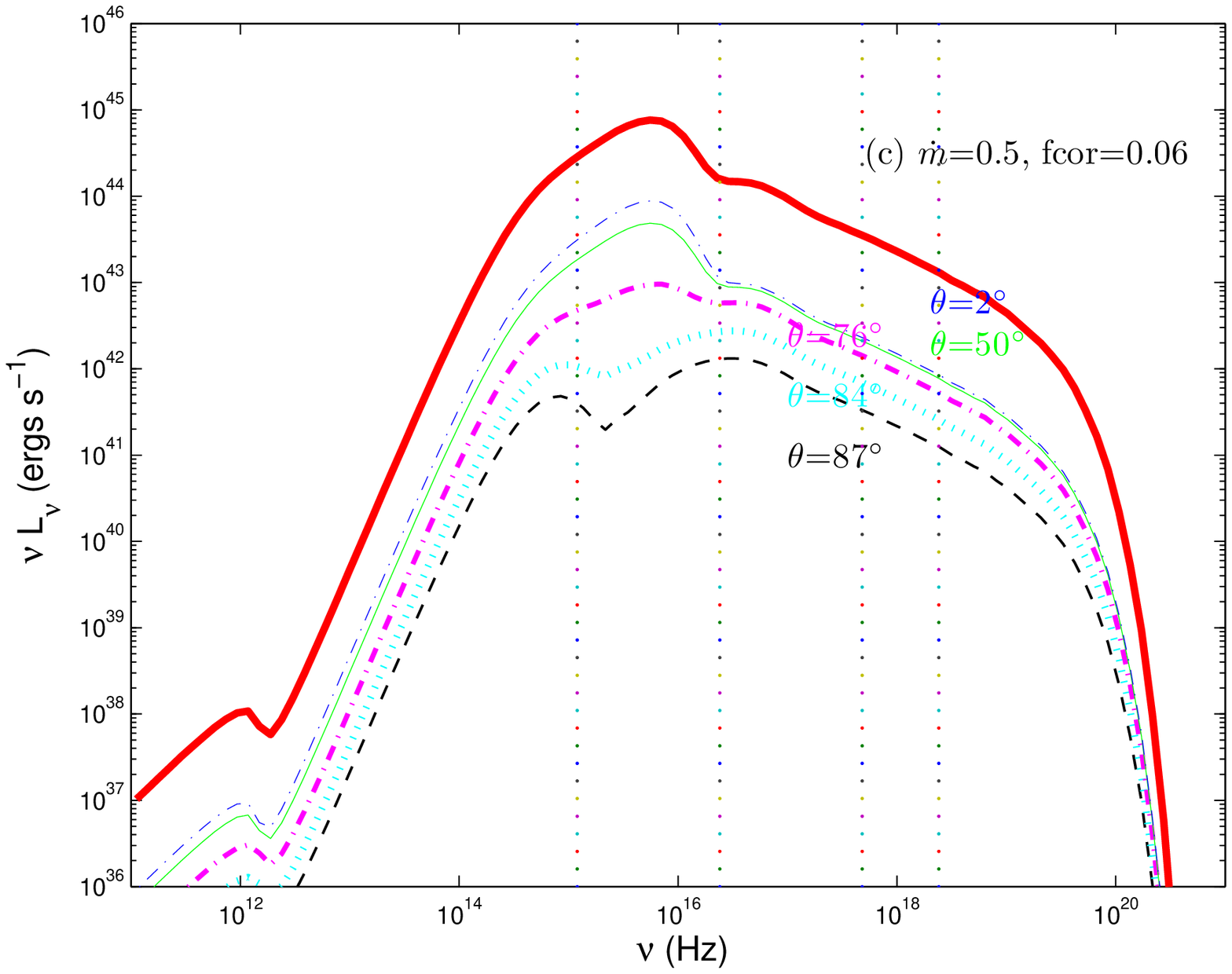,width=7.0cm,height=5.0
cm}}
\centerline{\psfig{figure=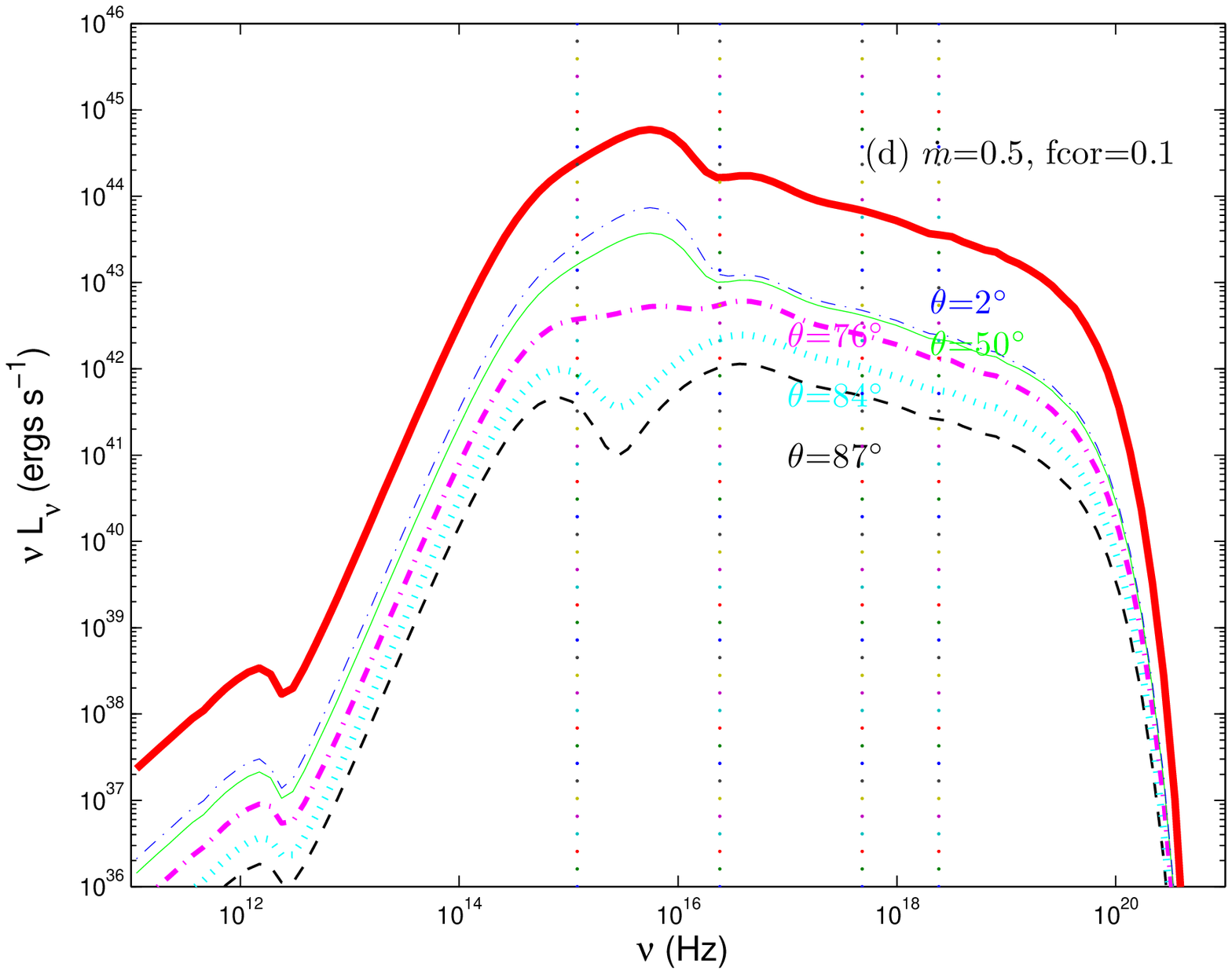,width=7.0cm,height=5.0
cm}} \caption{Spectra of the four disc-corona accretion models with
different values of parameters $\dot{m}$ and $f_{\rm cor}$ in four
plots, model (a): $\dot{m}=0.1$ and $f_{\rm cor}=0.1$, model (b):
$\dot{m}=0.1$ and $f_{\rm cor}=0.3$, model (c): $\dot{m}=0.5$ and
$f_{\rm cor}=0.06$, and model (d): $\dot{m}=0.5$ and $f_{\rm
cor}=0.1$. For each model, we plot the spectra observed at different
angles (the dashed lines in different colors). The emergent spectra
viewed at $\theta=87^{\circ}, 84^{\circ},76^{\circ}, 50^{\circ}$,
and $2^{\circ}$, are shown with black dashed, cyan dotted, magenta
dash-dotted, green thin solid, and blue thin dash-dotted lines,
respectively. The corresponding value of $\theta$ is denoted near
each spectrum. The red thick solid lines are the spectra integrated
over all directions ($\mu=0-1$). The four vertical dotted lines
represent the four typical frequencies corresponding to $2500 \rm
\AA$, 0.1 keV, 2 keV, and 10 keV, respectively. } \label{fig1}
\end{figure}


\begin{figure}
\centerline{\psfig{figure=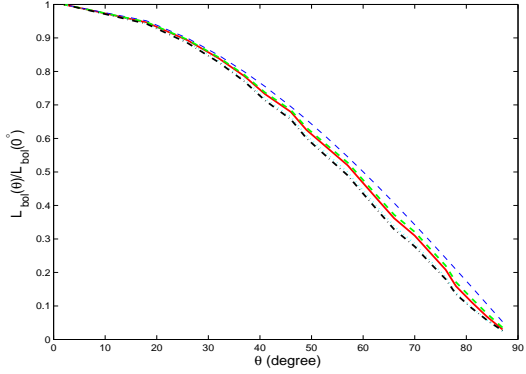,width=7.0cm,height=5.0
cm}}
 \caption{The observed bolometric luminosity varies with the viewing angle.
 The red solid, green dashed, cyan dotted, and black dash-dotted
lines correspond to the four models (a)-(d) respectively. The blue
thin dashed line indicates the relation of $L_{{\rm
bol}}\propto\cos\theta$. } \label{fig2}
\end{figure}


\begin{figure}
\centerline{\psfig{figure=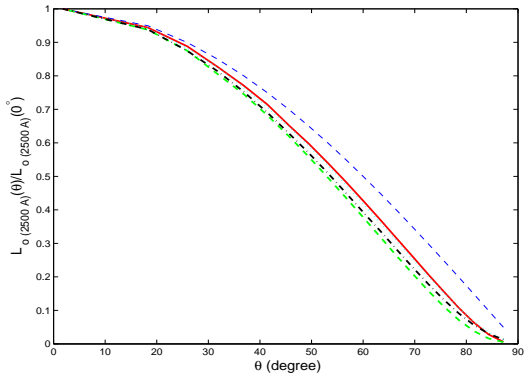,width=7.0cm,height=5.0
cm}}
 \caption{The spectral luminosity at optical/UV
band($\lambda=2500{\rm \AA}$) varies with the viewing angle. The red
solid, green dashed, cyan dotted, and black dash-dotted lines
correspond to the four models (a)-(d) respectively. The blue thin
dashed line represents the relation of $L_{\rm o}\propto\cos\theta$.
} \label{fig3}
\end{figure}


\begin{figure}
\centerline{\psfig{figure=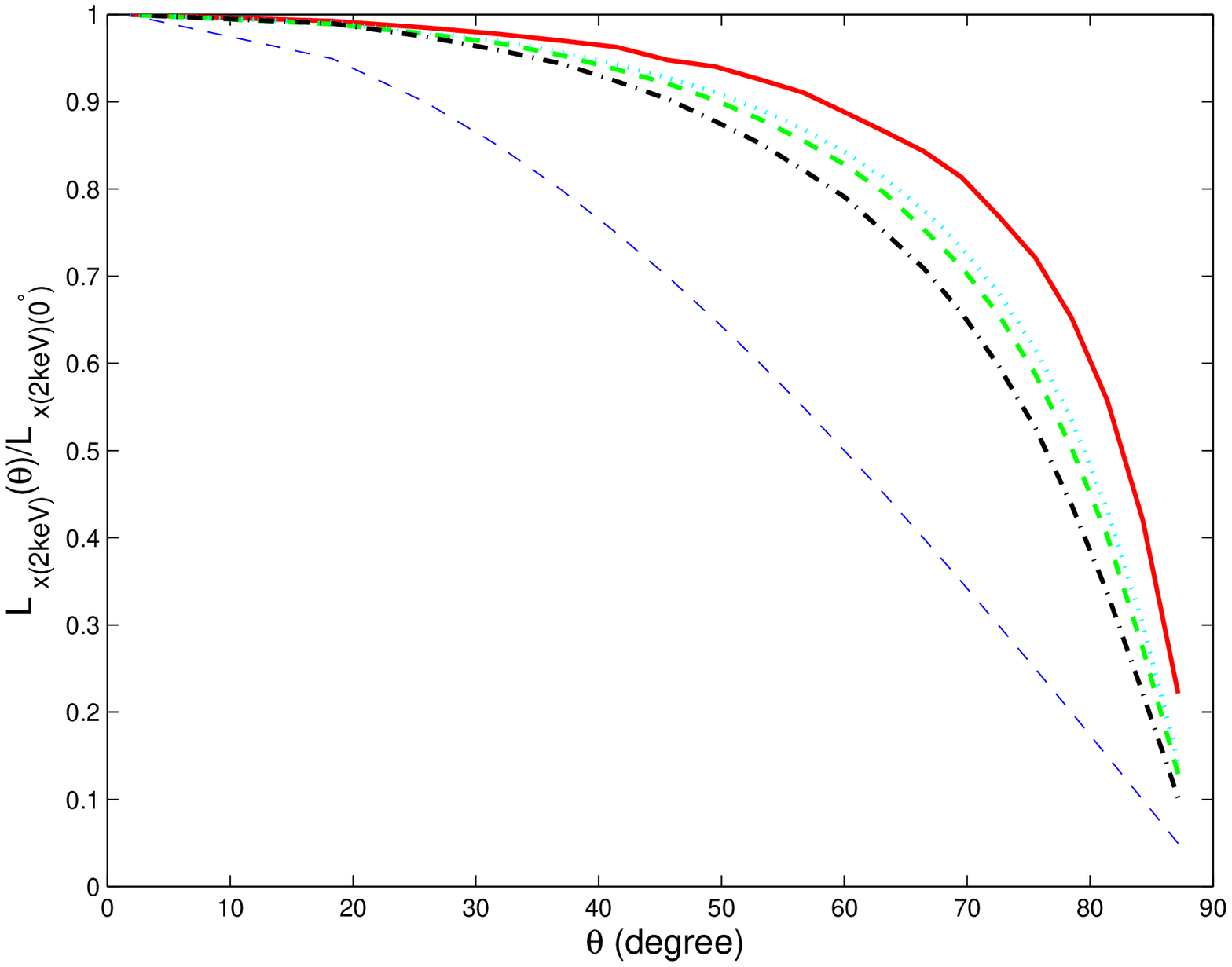,width=7.0cm,height=5.0
cm}}
 \caption{The spectral luminosity at X-ray band($E=2$
keV) varies with the viewing angle. The red solid, green dashed,
cyan dotted, and black dash-dotted lines correspond to the four
models (a)-(d) respectively. The blue thin dashed line represents
the relation of $L_{\rm x}\propto\cos \theta$. } \label{fig4}
\end{figure}


\begin{figure}
\centerline{\psfig{figure=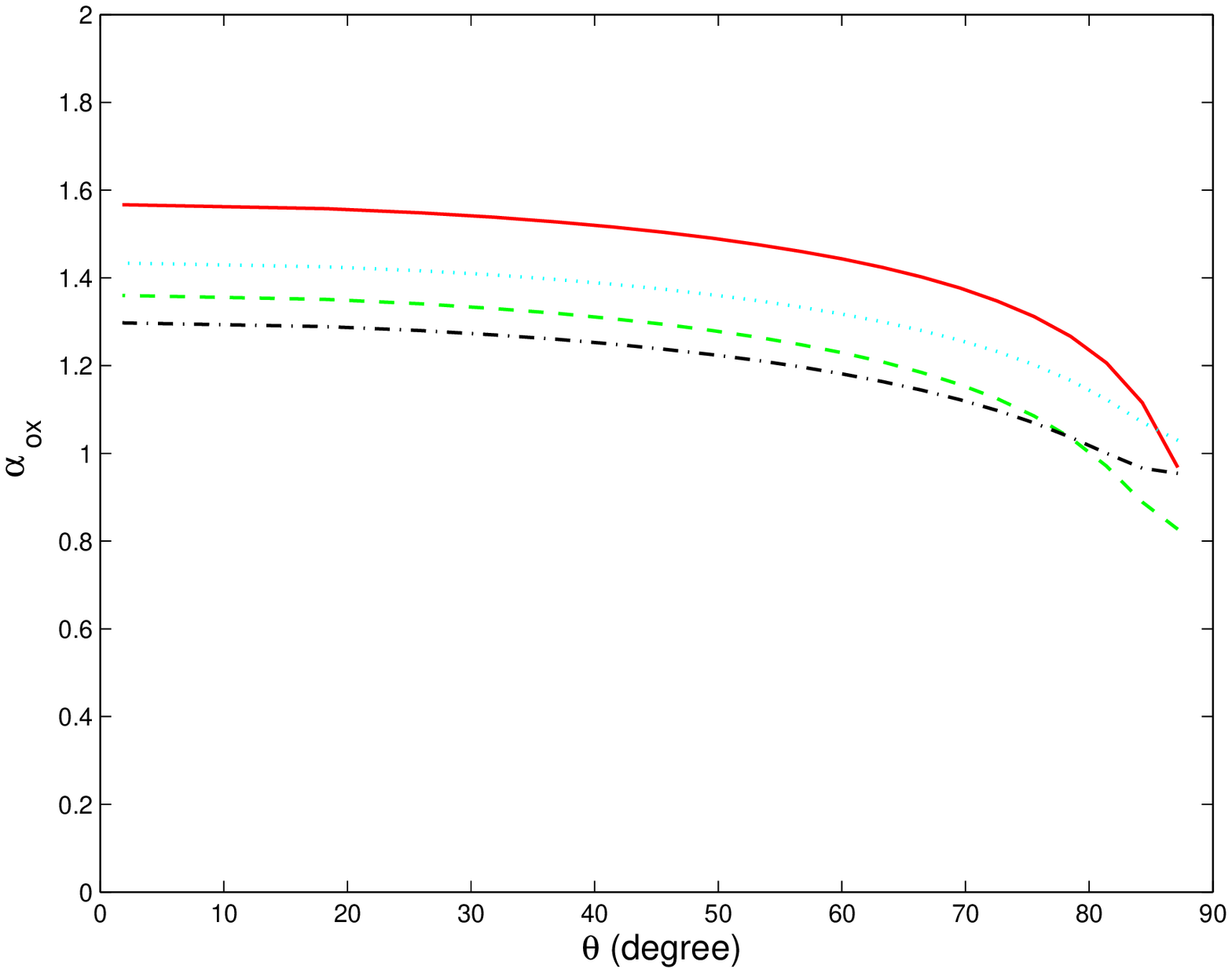,width=7.0cm,height=5.0
cm}}
\centerline{\psfig{figure=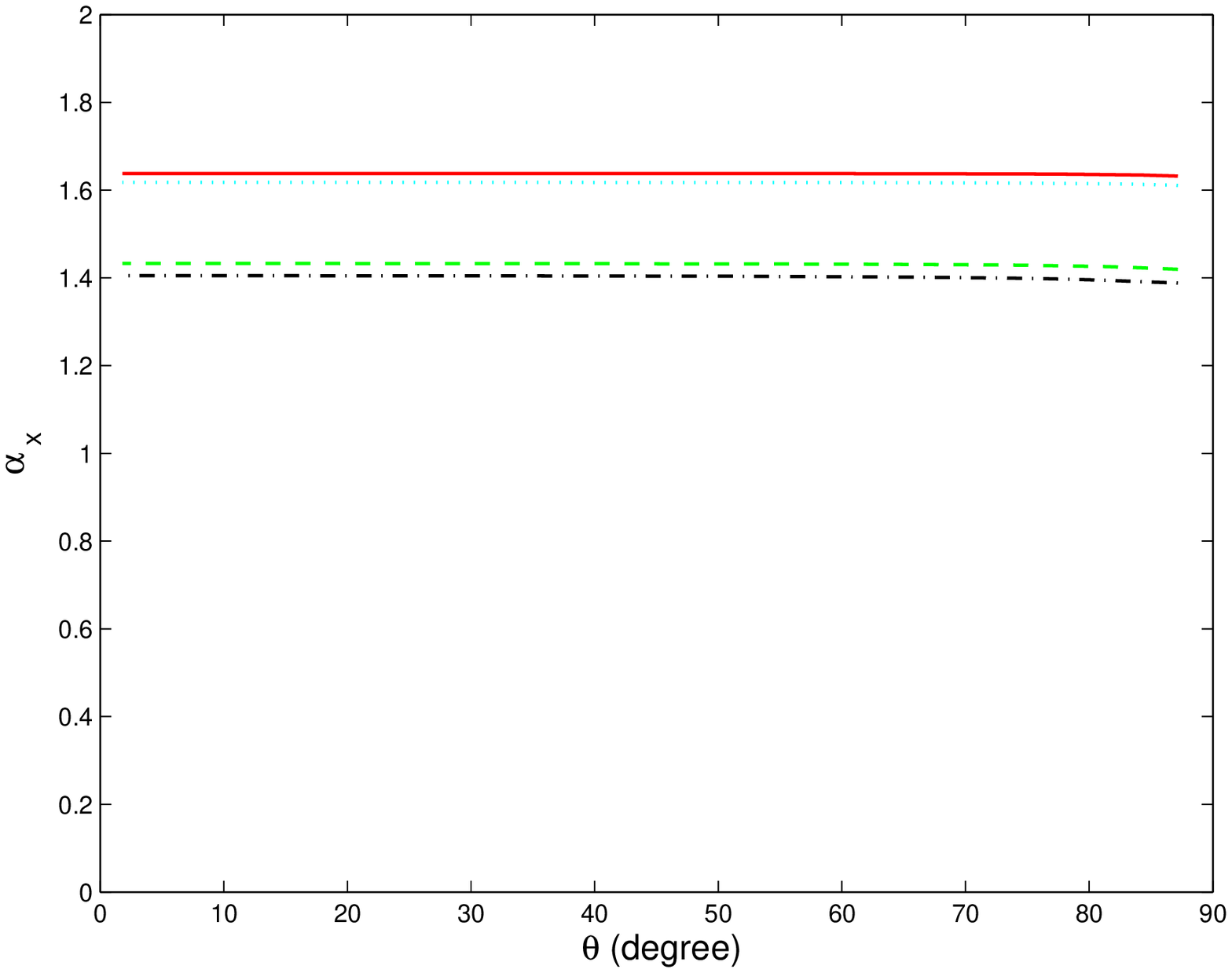,width=7.0cm,height=5.0
cm}}
 \caption{The optical/UV to X-ray spectral index,
$\alpha_{\rm ox}$(upper panel), and the X-ray spectral index between
2 keV and 10 keV, $\alpha_{\rm x}$(lower panel), vary with the
viewing angle. The red solid, green dashed, cyan dotted, and black
dash-dotted lines correspond to the four models (a)-(d)
respectively.} \label{fig5}
\end{figure}


\begin{figure}
\centerline{\psfig{figure=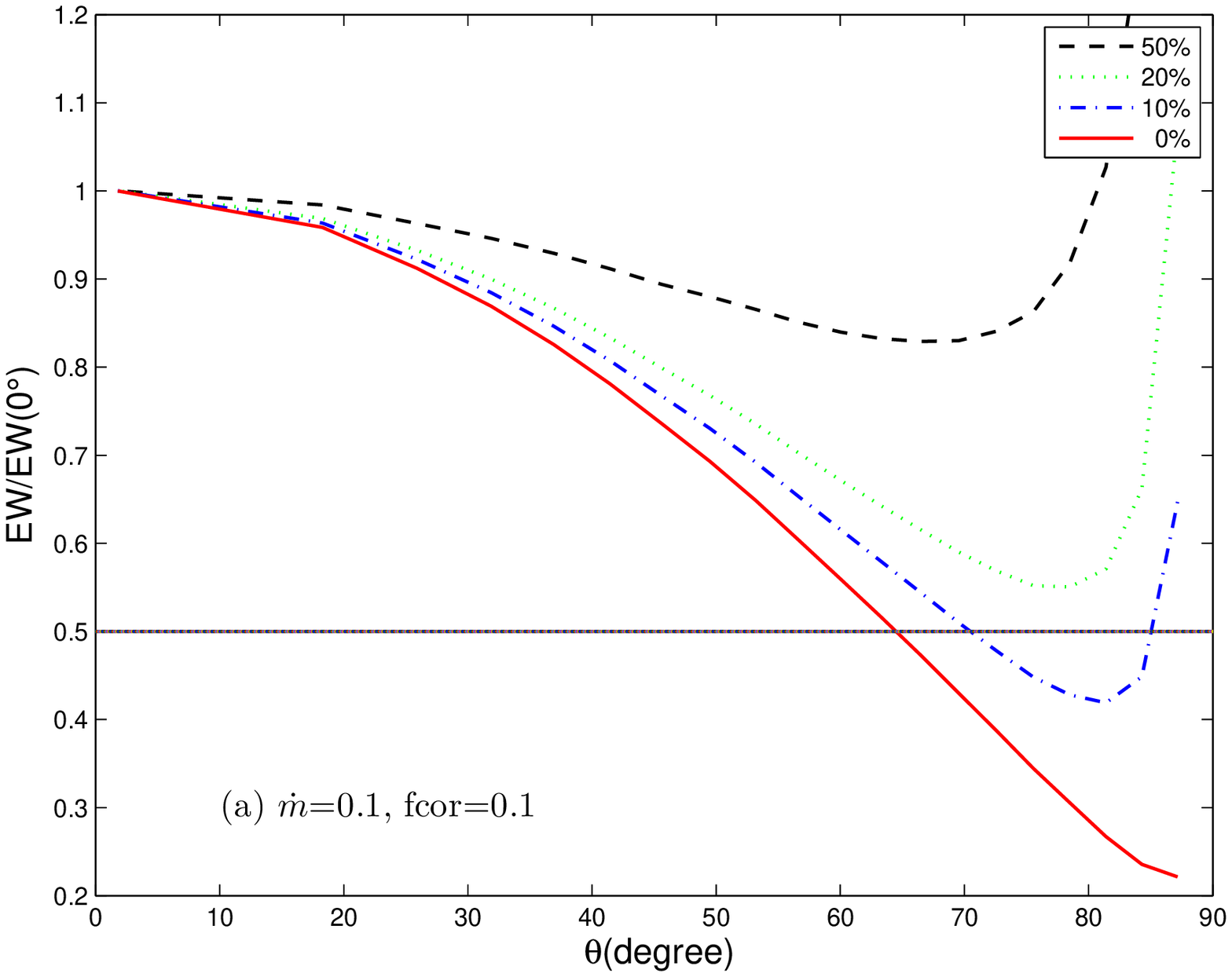,width=7.0cm,height=5.0
cm}}
 \caption{The equivalent width of the narrow Fe ${\rm
K}{\alpha}$ line emitted from AGN varies with the viewing angle,
which is calculated with model (a). The red solid line represents
the case that all the line emission is radiated from the thin
accretion disc. The black dashed, green dotted, and blue dash-dotted
lines are the results of the cases that 50$\%$, 20$\%$, and 10$\%$
of the line luminosity contributed from the torus, respectively. The
black dotted horizontal line represents the relative equivalent
width being $0.5$.} \label{fig6}
\end{figure}

\section{Discussion}

Solving the radiative transfer equation of the corona numerically,
we obtain the cooling rate in the hot corona and the emergent
spectrum of the accretion disc-corona system viewed at any specified
angle. The calculations are carried out for four sets of model
parameters, in which different accretion rate $\dot{m}$ and
different ratio of the power dissipated in the corona $f_{\rm cor}$
are adopted, but all the other model parameters are fixed. Overall,
the calculated spectra are sensitive to the values of the model
parameters. For the models with same value of $f_{\rm cor}$ [compare
Fig. 1(a) with 1(d)], the larger the accretion rate, the harder the
spectra from the disc-corona accretion flow are. The X-ray emission
predominantly originates from the inverse Compton scattering of the
soft photons from the thin disc by the hot electrons in the corona.
The larger the accretion rate, the more the soft photons are
radiated from the disc, thus the emission in the lower energy band
(i.e. optical/UV band) becomes relatively weaker while the high
energy band (i.e. X-ray band) becomes relatively stronger, and the
spectra are harder (see $\alpha_{\rm ox}$ in the left panel of Fig.
5). On the other hand, for the models with same value of $\dot{m}$
[compare Fig. 1(a) with 1(b), or Fig. 1(c) with 1(d)], the higher
the ratio of the power dissipated in the corona, $f_{\rm cor}$, the
harder the spectra from the disc-corona accretion flow are. We find
that the electron temperature of the corona increases with $f_{\rm
cor}$, which causes the Compton emission to be hard for the high
$f_{\rm cor}$ cases.

It is found that the spectra of disc-corona system observed at
different directions have the similar shapes in 2-10 keV X-ray band.
The X-ray spectral index, $\alpha_{\rm x}$, remains almost unchanged
for different viewing angle $\theta$ (see the lower panel of Figure
5). The X-ray emission mostly originates from the inverse Compton
scattering of the soft photons radiated from the thin disc by the
hot electrons in the corona. The inverse Compton scattered X-ray
spectrum mainly depends on spectrum of the seed photons and the
temperature of the hot electrons. Thus, the X-ray spectral index is
not sensitive to the viewing direction. Comparing the values of
$\alpha_{\rm x}$ for different models, we find that $\alpha_{\rm
x}\approx 1.64$ for the case with $\dot{m}=0.1$ and $f_{\rm
cor}=0.1$, while $\alpha_{\rm x}\approx 1.40$ for the case with
$\dot{m}=0.5$ and $f_{\rm cor}=0.1$. This result seems inconsistent
with the observed results and theoretical predictions in the
previous works which claim the increasing $\alpha_{\rm x}$ with the
Eddington ratio, $L_{\rm bol}/L_{\rm Edd}$ or accretion rate,
$\dot{m}$ \citep*[e.g., see Fig. 4 in][]{2009MNRAS.394..207C}. The
reason is that we employ the same value of parameter, $f_{\rm
cor}=0.1$, for these two models, which is inconsistent with the fact
that the value of $f_{\rm cor}$ decreases with the accretion rate
\citep*[e.g., see Fig. 1 in][]{2009MNRAS.394..207C}. If we compare
the results of the models (b) with (c), which adopt $f_{\rm
cor}=0.3$ for the case with $\dot{m}=0.1$ and a smaller $f_{\rm
cor}=0.06$ for the case with $\dot{m}=0.5$, we have $\alpha_{\rm
x}\approx 1.43$ for the case with $\dot{m}=0.1$ and $f_{\rm
cor}=0.3$ (see green line Figure 5), while $\alpha_{\rm x}\approx
1.62$ for the case with $\dot{m}=0.5$ and $f_{\rm cor}=0.06$ (see
cyan line in Figure 5).

The main focus of this work is to explore the change of the emergent
spectra from the disc-corona system viewed at different angles. In
Figure 1, we find that the luminosity decreases with the increasing
viewing angle $\theta$ with respect to the axis of the accretion
disc. The detailed results of the observed bolometric luminosity are
plotted in Figure 2, which shows that the bolometric luminosity is
nearly proportional to $\mu$ ($\mu$=cos$\theta$). This is due to the
area-projection effect. The change of the spectral shapes for
different viewing angles is different between optical/UV and soft
X-ray bands (see Figure 1). The observed spectral luminosity at a
typical optical/UV band of $2500 {\rm \AA}$ decreases with $\theta$
(see Figure 3). It decreases more rapidly than
$\cos\theta$-relation. The blackbody emission from the thin disc is
partly absorbed in the corona. For the optical/UV emission from the
disc, the specific intensity of the photons from the upper surface
of the corona, $I_{\nu}\sim I_{\nu,0}\exp({-\tau_{0}/\mu})$, where
$I_{\nu,0}$ is the specific intensity of the photons injected in the
lower surface of the corona, and $\tau_{0}$ is the vertical optical
depth of the corona. Thus, the observed spectral luminosity at
optical/UV band is proportion to $\mu\exp({-\tau_{0}/\mu})$, which
decreases more quickly with $\theta$ than $\cos\theta$-relation. On
the other hand, although the X-ray spectral shape remains unchanged
for different $\theta$, the observed spectral luminosity at 2 keV
decreases with $\theta$ (see Figure 4). It decreases more slowly
than $\cos\theta$-relation.

The narrow 6.4 KeV Fe K$\alpha$ lines are ubiquity in AGNs. Although
its origin is still uncertain, it was suggested that they may
probably originate from the distant molecular cloud (torus), the
outer accretion disc or/and the broad-line region (BLR). The BLR
origin is ruled out by the fact that no correlation between the Fe
K$\alpha$ core width and the BLR line (i.e., H$\beta$) width
\citep{2006MNRAS.368L..62N}. The emitted narrow 6.4 KeV line are
found to show different properties for different types of AGN. The
line luminosities of the narrow 6.4 KeV Fe K$\alpha$ line from type
${\rm \uppercase\expandafter{\romannumeral 1}}$ AGNs are much
stronger than those from type ${\rm
\uppercase\expandafter{\romannumeral 2}}$ AGNs at the same X-ray
continuum luminosity. Compiling 89 Seyfert galaxies and using [O IV]
emission to estimate the intrinsic luminosity of the sources,
\citet{2010ApJ...725.2381L} found that the Fe K$\alpha$ line
luminosities of Compton-thin Seyfert 2 galaxies are in average 2.9
times weaker than their Seyfert 1 counterparts.
\citet{2014MNRAS.441.3622R} found that the Fe K$\alpha$ line
luminosity is correlated with the 10-50~keV X-ray continuum
luminosity either for Sy1s and Sy2s. The slopes of the correlations
are almost same for these two types of the sources, but the Fe
K$\alpha$ line luminosities of Sy1s are about twice of those for
Sy2s at a given X-ray continuum luminosity.

It is believed that type 1 Seyfert galaxies are intrinsically same
as type 2 Seyferts but viewed at different angles
\citep*[][]{1993ARA&A..31..473A}. We find that the observed
systematical difference of EW of Fe K$\alpha$ emission lines between
Sy1s and Sy2s can be attributed to the difference of the view
angles(see Figure 6). Such EW difference can be reproduced by our
model calculations, provided Sy1s are observed in nearly face-on
direction and the average inclination angle of Sy2s $\sim 65^\circ$,
which support the unification scheme of AGN. If a fraction of Fe
line emission is contributed by the torus, a larger average
inclination angle is required for Sy2s, which implies that the
contribution from the torus should be much less than that from the
disc.

No significant correlation is found between the spectral index
$\alpha_{\rm ox}$ and the radio core dominance parameter $R$, which
is believed to be an indicator of the viewing angle
\citep*[][]{2013MNRAS.435.3251R}. Our results show that the spectral
index $\alpha_{\rm ox}$ is weakly dependent on the inclination angle
(see Figure 5). This is not surprising, because there is a strong
correlation between $\alpha_{\rm ox}$ and optical
luminosity/Eddington ratio
\citep*[e.g.,][]{2003AJ....125..433V,2010ApJS..187...64G,2010A&A...512A..34L}.
This correlation may smear out any possible correlation between
$\alpha_{\rm ox}$ and inclination angle for a normal AGN sample. We
suggest that the investigation on $\alpha_{\rm ox}$ and inclination
angle relation should be carried out with a sample of AGNs in the
narrow range of Eddington ratio/luminosity.

In the standard unification model of AGNs, the different observation
properties of different types of AGN, such as, different continuum
spectral shape and broad emission lines, can be explained by their
different inclination angles of the black hole accretion flow and
the surrounding torus. We show in this work that the emission
spectra emitted from the accretion flow of a black hole are
anisotropic along the different viewing angles, which is consistent
with the area-projection effect in the bolometric luminosity, but
very different in the other characteristics of SED, i.e., the
spectral luminosities in the optical/UV band and the X-ray band. Our
calculations of accretion disc-corona spectra may provide more
precise orientation effect correction in predicting the intrinsic
luminosity functions of AGN sources from the observed luminosities
at certain band. Recently, the effect of anistropic radiation from
the accretion discs on the luminosity function derived from an AGN
sample was evaluated by \citet{2014ApJ...787...73D}, and they found
that the bright end of the luminosity function may be overestimated
by a factor of $\sim 2$ without considering this effect. A simple
$\cos\theta$-dependent specific intensity from a bare accretion disc
is used in their estimates. Our present detailed calculations of the
disc corona spectra as functions of the viewing angle can be
incorporated in deriving the intrinsic AGN luminosity.

The calculations of the radiation transfer in the corona in this
work are carried out in Newtonian frame. A general relativistic
accretion corona model is required for direct modeling the observed
spectra of AGN. The calculations in this work can be easily expanded
for the accretion discs surrounding Kerr black holes in general
relativistic frame. This will be reported in our future work.

\section*{Acknowledgments}

We thank the referee for his/her helpful comments. This work is
supported by the NSFC (grants 11078014, 11233006 and 11220101002).

\end{document}